\documentclass[twocolumn,showpacs,preprintnumbers,amsmath,amssymb]{revtex4}

\usepackage{graphicx}
\usepackage{dcolumn}
\usepackage{bm}

\begin{document}

\preprint{APS/123-QED}

\title{Spin and Valley Splittings in Multilayered Massless Dirac Fermion System}

\author{Naoya Tajima$^1$}
\author{Mitsuyuki Sato$^2$}%
\author{Shigeharu Sugawara$^3$}%
\author{Reizo Kato$^1$}%
\author{Yutaka Nishio$^2$}%
\author{Koji Kajita$^2$}%

\affiliation{%
$^1$RIKEN, Hirosawa 2-1, Wako-shi, Saitama 351-0198, Japan \\
$^2$Department of Physics, Toho University, Miyama 2-2-1, Funabashi-shi, Chiba 274-8510, Japan \\
$^3$Department of Physics, Faculty of Science and Technology, Tokyo University of Science, Yamazaki 2641, Noda, Chiba 278-8510, Japan
}%

\date{\today}

\begin{abstract}
The inter-layer magnetoresistance in a multilayered massless Dirac fermion system, $\alpha$-(BEDT-TTF)$_2$I$_3$, under hydrostatic pressure was investigated. We succeeded in detecting the zero-mode (n=0) Landau level and its spin splitting in the magnetic field normal to the 2D plane. We demonstrated that the effective Coulomb interaction in the magnetic field intensifies the spin splitting of zero-mode Landau carriers. At temperatures below 2K, magnetic fields above several Tesla break the twofold valley degeneracy.
\end{abstract}


\maketitle

Since Novoselov {\it et al.} \cite{rf:1} and Zhang {\it et al.} \cite{rf:2} experimentally demonstrated that graphene is a zero-gap system with massless Dirac particles, such systems have fascinated physicists as a source of exotic systems and/or new physics. At the same time, the quasi-two-dimensional (Q2D) organic conductor $\alpha$-(BEDT-TTF)$_2$I$_3$ (BEDT-TTF=bis(ethylenedithio) tetrathiafulvalene) \cite{rf:3} was found to be a new type of massless Dirac fermion (MDF) system under high pressure \cite{rf:4, rf:5, rf:6, rf:7, rf:8, rf:9}. In contrast to graphene, this is the first bulk (multilayered) zero-gap material with Dirac cone type energy dispersion. Another important difference is the fact that according to the band calculation \cite{rf:4, rf:5} or the first principles band calculation \cite{rf:6}, the present system has tilted Dirac cones. Hence, the Fermi velocity, $v_{\rm F}$, is highly anisotropic. Thus, $\alpha$-(BEDT-TTF)$_2$I$_3$ provides us a new type of MDF system with a layered structure and an anisotropic $v_{\rm F}$. 

One of the characteristic features of transport in the multilayered MDF system is clearly observed in the inter-layer magnetoresistance shown in Fig. 1(a) \cite{rf:9}. The results are interpreted as follows.

In the magnetic field, the energy of Landau levels (LLs) in zero-gap systems is expressed as $E_{\rm nLL}=\pm \sqrt{2e \hbar v_{\rm F}^2 |{\rm n}||B|}$, where n is the Landau index and $B$ is the magnetic field strength \cite{rf:10}. One important difference between zero-gap conductors and conventional conductors is the appearance of a (n=0) LL at zero energy when magnetic fields are applied normal to the 2D plane. This special LL is called the zero-mode. Since the energy of this level is $E_{\rm F}$ irrespective of the field strength, the Fermi distribution function is always 1/2. It means that half of the Landau states in the zero mode are occupied. Note that in each LL, there are states whose density is proportional to $B$. The magnetic field, thus, creates mobile carriers. 

For $k_{\rm B}T < E_{\rm 1LL}$, most of the mobile carriers are in the zero-mode. Such a situation is called the quantum limit. The carrier density per valley and per spin direction in the quantum limit is given by $D(B)=B/2\phi_{0}$, where $\phi_{0}=h/e$ is the quantum flux. The factor 1/2 is the Fermi distribution function at $E_{\rm F}$. This effect is detected in the inter-layer resistance, $R_{zz}$, in the longitudinal magnetic field. In this field configuration, the interaction between the electrical current and the magnetic field is weak because they are parallel to each other. Hence, the effect of the magnetic field appears only through the change in the carrier density. In this regard, the large change in the density of zero-mode carriers gives rise to remarkable negative magnetoresistance in the low magnetic field region, as shown in Fig. 1(a). It is written as $R_{zz} \propto 1/B$. This behavior agrees well with the realistic theory of Osada \cite{rf:11}. 

At a high magnetic field, on the other hand, the Zeeman effect plays an important role in zero-mode carriers. In a magnetic field, each LL is split into two levels with energies $E_{\rm nLL}\pm \Delta E_{\rm s}/2$, where $\Delta E_{\rm s}$=$g^*\mu_{\rm B}B$ is the spin-splitting and $g^*$ is the effective g-factor. This change in the energy structure gives rise to a change in the carrier density in LLs. At a low temperature and/or a high magnetic field, where $k_{\rm B}T < \Delta E_{\rm s}$, this effect becomes important because the energy level is shifted from the position of $E_{\rm F}$. It works to reduce the density. The magnetoresistance changes from negative to positive, and is dependent on the magnetic field and the temperature as 
\begin{equation}
R_{zz} \propto \displaystyle {1}/{D(B)}\cdot \exp({g^*\mu_{\rm B}B}/{2k_{\rm B}T}).
\label{eqn:1}
\end{equation}
Hence, $R_{\rm zz}\cdot B$ obeys the exponential law as shown in Fig. 1(b) at a high magnetic field. 

The detection of the zero-mode and its spin-split level at a low field strongly indicates the high purity of our system and leads our investigation to a new stage. Each zero-mode in this system has twofold degeneracy (valley degeneracy) originating from two Dirac cones in the first Brillouin zone \cite{rf:4, rf:5, rf:6}. According to the simple Storner-like theory of quantum Hall ferromagnetism \cite{rf:12}, the Coulomb interaction plays an important role in spin and valley symmetries in the magnetic field. In this paper, we experimentally demonstrate that the valley degeneracy in this system is broken at magnetic fields above several Tesla at temperatures below 2 K. In graphene, on the other hand, this state was realized in the magnetic field above 20 T \cite{rf:13}. This critical field strength depends on the scattering broadening energy of the zero-mode. The broadening energy of the LLs in our system is much lower than that in graphene. Another important fact is that according to the theory of Kobayashi {\it et al.}, when the Dirac cone is tilted, the long-range Coulomb interaction has finite matrix elements between valleys, and it breaks the valley degeneracy in the magnetic field (Fig. 2) \cite{rf:14}.

We succeeded in obtaining evidence of the valley-split levels in our material, as follows. 

In the magnetic field, the effective Coulomb interaction enhances the spin splitting as, 
\begin{equation}
\Delta E_{\rm s} = g\mu_{\rm B}B +\sqrt{\pi / 2}I,
\label{eqn:2}
\end{equation}
where $g \sim$ 2 is the g-factor, $I \sim e^2/\epsilon l_{\rm H} \sim 50 \sqrt{B}/\epsilon$ meV is the effective Coulomb interaction in the magnetic field, $l_{\rm H}=\sqrt{\hbar/eB}$ is the magnetic length, and $\epsilon$ is the effective polarizability. Then, we have the effective g-factor,
\begin{equation}
g^*=\displaystyle {\Delta E_{\rm s}}/{\mu_{\rm B}B}=g+{\sqrt{\pi / 2}I}/{\mu_{\rm B}B}.
\label{eqn:3}
\end{equation}
It should exceed 2 and depends on the magnetic field because $I\propto \sqrt{B}$. 

This effect can be seen in $R_{zz}$ under a longitudinal magnetic field (Fig. 1(a)). At the minimum, we have $g^* \mu_{\rm B}B \sim \Gamma$, where $\Gamma$ is the scattering broadening energy or the thermal broadening energy of each LL \cite{rf:11}. At a high magnetic field, on the other hand, $R_{zz}$ obeys an exponential law as Eq. (\ref{eqn:1}). Note that $R_{zz}$ can detect only the charge gap, $\Delta E$, between levels that are nearest to $E_{\rm F}$, assuming that $E_{\rm F}$ always locates at the contact point. In the valley degenerate state, hence, $\Delta E = \Delta E_{\rm s}=g^* \mu_{\rm B} B$. The magnetoresistance in the valley splitting state, however, is characterized by the charge gap between $(\downarrow, -)$ and $(\uparrow, +)$, $\Delta E = \Delta E_{\rm s}-\Delta_{\rm v}$, where $\Delta_{\rm v}$ is the valley splitting, and we describe the level structure of the zero mode using spin index $(\uparrow, \downarrow)$ and valley index $(+, -)$, as shown in Fig. 2. 

In our previous work, we published data for $R_{zz}$ at magnetic fields of up to 7 T and succeeded in detecting the zero-mode and its spin-split level \cite{rf:9}. To demonstrate that the magnetic fields break the valley degeneracy in this system at low temperatures, however, a higher magnetic field becomes important. In this work, we examined the magnetoresistance at magnetic fields of up to 14 T in the temperature region below 4.1 K. 

A sample to which four electrical leads were attached was encased in a Teflon capsule filled with pressure medium (Idemitsu DN-oil 7373). The capsule was set in a clamp-type pressure cell made of MP35N hard alloy and hydrostatic pressure of up to 1.7 GPa was applied. Resistance measurements were carried out using the conventional DC method with the electrical current applied along the $c$-crystal axis, which is normal to the 2D plane.


\begin{figure}
\includegraphics[viewport = 20 150 550 680, scale=.35, clip]{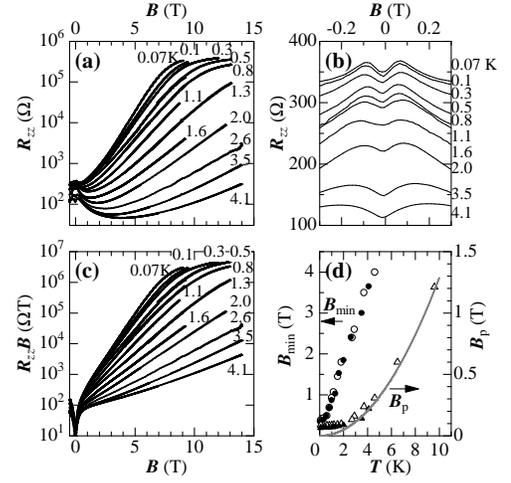}
\caption{\label{fig:1} (a) Magnetic field dependence of $R_{zz}$ below 4.1 K under pressure of approximately 1.7 GPa. (b) $R_{zz}$ in the low field region below 0.4 T. (c) Magnetic field dependence of $R_{zz} \cdot B$. (d) Temperature dependence of $B_{\rm p}$ (solid triangles) and $B_{\rm min}$ (solid circles). We show $B_{\rm p}$ and $B_{\rm min}$ estimated from previous data \cite{rf:9} by open triangles and open circles. The solid line is the curve of $E_{\rm 1LL}$ with $v_{\rm F} \sim 4\times 10^4 {\rm m/s}$. } 
\end{figure}

\begin{figure}
\includegraphics[viewport = 20 380 550 620, scale=.35, clip]{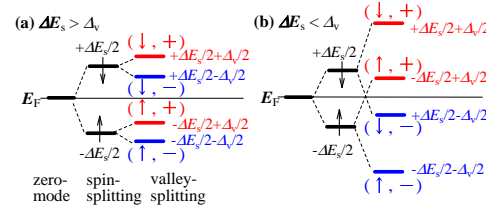}
\caption{\label{fig:2} (color online). Zero-mode and its spin- and valley-split levels for the cases $\Delta E_{\rm s}>\Delta_{\rm v}$ (a) and $\Delta E_{\rm s}<\Delta_{\rm v}$ (b). The level structure is described using spin index $(\uparrow, \downarrow)$ and valley index $(+, -)$}.
\end{figure}

Figure 1(a) shows the magnetic field dependence of $R_{zz}$ below 4.1 K. It reproduces well our previous data published in Ref. \cite{rf:9}. We can understand this magnetoresistance in terms of zero-mode carriers, including the spin splitting. In the interpretation of $R_{zz}$ in Ref. \cite{rf:9}, however, we did not take into consideration the Coulomb interaction. The Coulomb interaction should be taken into consideration because it has a significant influence on the transport phenomena in the magnetic field. 

Thus, the first step is to verify the effective Coulomb interaction in this system from the effective g-factor $g^*_{\rm min}$ estimated from the relationship $g^*_{\rm min} \mu_{\rm B} B_{\rm min} \sim \Gamma$ at the magnetoresistance minimum. Here, broadening energy $\Gamma$ should be proportional to $k_{\rm B}T$ because $B_{\rm min}$ decreases linearly with decreasing temperature, as shown in Fig. 1(d). It can be written approximately as $\Gamma = \Gamma_0 + 2 k_{\rm B}T$ from the investigation of the relationship between $\Gamma$ and the temperature from the simulation calculation of $R_{zz}$ along Ref. \cite{rf:11}. The scattering broadening energy, $\Gamma_0$, is roughly estimated to be about 3 K as follows.

Since each LL is broadened by the scattering of carriers and/or thermal energy, the zero-mode is sure to overlap with the other LLs at a low magnetic field. In such a region, the relationship of $R_{zz} \propto 1/ B$ loses its validity. We can recognize this region in Fig. 1(b) where a positive magnetoresistance is observed. This critical field, $B_{\rm p}$, shifts to a lower field with decreasing temperature down to about 2 K, where it almost saturates at about 0.04 T, as shown in Figs. 1(b) and (d). The overlap between the zero-mode and other LLs, primarily the n=1 LL will be sufficiently small above $B_{\rm p}$ and as a result, the negative magnetoresistance is observed there. Then, we have a tentative relationship: $E_{\rm 1LL} \sim 2k_{\rm B}T_{\rm p}$ at $B_{\rm p}$. In fact, $E_{\rm 1LL}$ with $v_{\rm F} \sim 4\times 10^4$ cm/s is reproduced well except in the temperature region below 2 K. This Fermi velocity $v_{\rm F}$ corresponds to that estimated from the temperature dependence of the carrier density, $N$, written as $N \propto (T / \bar{v}_{\rm F})^2$, with $\bar{v}_{\rm F} \sim 10^5$ m/s within a factor of 3 \cite{rf:8}. The discrepancy of the data from the curve of $E_{\rm 1LL}$ below 2 K, on the other hand, suggests that thermal energy is sufficiently lower than the scattering broadening energy $\Gamma_0$. Thus, $\Gamma_0$ is roughly estimated to be approximately 3 K from the constant value of $B_{\rm p}$ as $\Gamma_0 = \sqrt{2e\hbar v_{\rm F}^2 |B_{\rm p}|}$ \cite{rf:15}. Note that in the estimation of $\Gamma_0$, we did not consider the spin splitting because $\Gamma_0$ has little impact. This scattering broadening energy is much lower than that of graphene. In graphene, $\Gamma_0$ was estimated to be about 30 K \cite{rf:16}. 

\begin{figure}
\includegraphics[viewport = 25 400 550 650, scale=.4, clip]{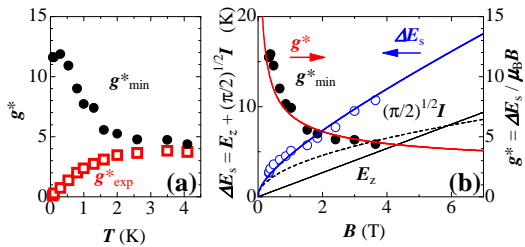}
\caption{\label{fig:3} (color online). (a) Temperature dependence of $g^*_{\rm min}$ (solid circles) and $g^*_{\rm exp}$ (squares). (b) Magnetic field dependence of $\Delta E_{\rm s}$ and $g^*_{\rm min}$. Each data can be reproduced well with Eq. (\ref{eqn:2}) (blue line) or (\ref{eqn:3}) (red line), with $\epsilon \sim$ 190. }
\end{figure}

Using the relationship $g^*_{\rm min} \mu_{\rm B} B_{\rm min} \sim \Gamma$, $g^*_{\rm min}$ at each temperature is estimated as shown in Figs. 3(a) and (b). It gives us definite evidence that the effective Coulomb interaction intensifies the spin splitting as follows.

First, at each temperature or magnetic field, $g^*_{\rm min}$ exceeds 2. It increases by about three times from about 4.5 at 4.1 K to nearly 12 at 0.1 K, as shown in Fig. 3(a). Note that this strong temperature dependence of $g^*_{\rm min}$ may arise from the effect of its strong magnetic field dependence because $B_{\rm min}$ depends strongly on temperature, as shown in Fig. 1(d). In fact, we can find this effect in the magnetic field dependence of $g^*_{\rm min}$ shown in Fig. 3(b). In the case of a system without the effective Coulomb interaction ($I$=0), $g^*$ should always be 2 in the magnetic field. However, $g^*_{\rm min}$ depends strongly on the magnetic field, as shown in Fig. 3(b). It obeys $1/\sqrt{B}$. Assuming that $\epsilon$ is independent of temperature and using Eq. (\ref{eqn:3}), with $\epsilon \sim$ 190, we try to fit the curve in Fig. 3(b). This simple formula reproduces the data well and shows evidence that the effective Coulomb interaction plays an important role in the spin splitting of this system. In particular, at a low magnetic field, this effect is strengthened. Here, we note that $\epsilon \sim$ 190 is the effective polarizability when the Coulomb interaction engages directly in the spin splitting. Thus, this is the upper limit of this system. Recently, Morinari and Tohyama performed a simulation calculation for in-plane magnetoresistance using this effective polarizability \cite{rf:17}, and our anomalous data \cite{rf:7} are quantitatively reproduced.

To conclude, we can detect the effect of Coulomb interaction probed by inter-layer magnetoresistance measurements in this system. According to the simple Storner-like theory of quantum Hall ferromagnetism \cite{rf:12}, the strong Coulomb interaction leads us to summarize that the twofold valley degeneracy may be broken at a high magnetic field that satisfies $\sqrt{\pi/2}I>\Gamma$. 

The last step is to determine the effect of valley splitting in the magnetic field. At 7 T, for example, the effective Coulomb energy exceeds $\Gamma$ below 2 K. Thus, we expect that this effect appears in the effective g-factor, $g^*_{\rm exp}$, estimated from the resistance curve obeying Eq. (\ref{eqn:1}) as shown in Fig. 1(b), at temperatures below 2 K and 7 T.

At temperatures above 2 K, $g^*_{\rm exp} \sim g^*_{\rm min}$, as shown in Fig. 3(a). The slight difference arise from the effective Coulomb interaction obeying Eq. (\ref{eqn:3}). At 7 T, for example, $\Delta E =g^*_{\rm exp} \mu_{\rm B}B \sim \Delta E_{\rm s} \sim$ 18 K (Fig. 4).

\begin{figure}
\includegraphics[viewport = 20 360 550 650, scale=.35, clip]{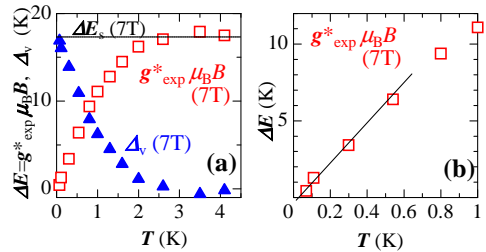}
\caption{\label{fig:4} (color online). (a) Temperature dependence of $\Delta E$ (squares) and $\Delta_{\rm v}$ (triangles) for $B$= 7 T. Solid line is $\Delta E_{\rm s} \sim$ 18 K prospected at 7 T. (b) $\Delta E$ for $B$= 7 T below 1 K.}
\end{figure}

The behavior of $g^*_{\rm exp}$ and $\Delta E$ below 2 K is much more impressive. Both abruptly drop by approximately 1/50 times from $g^*_{\rm exp} \sim$ 3.8 and $\Delta E \sim$ 18 K at 2 K to $g^*_{\rm exp} \sim$ 0.08 and $\Delta E \sim$ 0.4 K at 0.07 K. Judging from the detection of a strong effective Coulomb interaction, we ascribe the drop of $g^*_{\rm exp}$ and $\Delta E$ in Figs. 3(a) and 4 to the breaking valley degeneracy in the magnetic field above several Tesla, as shown in Fig. 2(a). In graphene, it was realized at magnetic fields above 20 T \cite{rf:12} and the realistic theory established that the strong Coulomb interaction broke the valley symmetry at a high magnetic field \cite{rf:13}. In our system, on the other hand, $\Gamma$ is much lower than that of graphene and therefore, we could detect the effect of valley splitting at a low magnetic field. At 2 K, for example, the critical magnetic field strength that satisfies $\sqrt{\pi/2}I > \Gamma$ is about 7 T. At 1 K, it is about 4 T.

Based on this speculation, valley splitting in our system can be estimated tentatively as $\Delta_{\rm v}=\Delta E_{\rm s}-\Delta E$. At 7 T, for example, valley splitting abruptly emerges at about 2 K and increases up to approximately 17 K at 0.07 K, as shown in Fig. 4(a). 

Recently, Kobayashi {\it et al.} predicted the pseudo-spin polarized ferromagnetic state with an easy plane where $\Delta_{\rm v} > \Delta E_{\rm s}$ as shown in Fig. 2(b) \cite{rf:14}. In this state, excited vortices and anti-vortices can move and the system may undergo the Kosterlitz-Thouless (KT) transition to form pairs of vortices and anti-vortices at the lowest temperature. Our data, however, suggest the situation $\Delta_{\rm v} < \Delta E_{\rm s}$, as shown in Fig. 2(a). 

Examining $\Delta E$ data below 1 K in more detail (Fig. 4(b)), we find that it seems to go across zero and then the pseudo-spin polarized ferromagnetic state in which the levels between $(\downarrow, -)$ and $(\uparrow, +)$ cross may be realized below 0.07 K (Fig. 2(b)). We also expect that this state would be realized at a high magnetic field. The data below 0.8 K in Figs. 1(a) and (b) are apparently discrepant from Eq. (\ref{eqn:1}) at a high magnetic field. The saturation of $R_{zz}$ may indicate a symptom of the crossover to the pseudo-spin polarized ferromagnetic state at a high magnetic field, as shown in Fig. 2(b). 

Lastly, we briefly mention the mobility edge in this system at a low temperature and a high magnetic field. In Fig. 4(b), we perceive the fact that $\Delta E$ below 0.1 K is lower than $\Gamma_0 \sim$ 3 K and yet, $R_{zz}$ at 7 T is much higher than that at the minimum. At 0.07 K, for example, $\Delta E$ is approximately 0.4 K. It arises from the localization of electrons on the tails of the broadened zero-mode. When LL is broadened by scattering, only the electrons at the vicinity of center $E_{\rm nLL}$ are mobile. This critical energy $E_{\rm nLL} \pm E_{\rm c}$ is called mobility edge. The smaller the Landau radius is, the stronger the localization is. The effect of localization also depends on the temperature. Assuming that the mobility edge in this system locates at the vicinity of the center of each Landau level, $E_{\rm c}$ should be less than $\Delta E/2 \sim$ 0.2 K at 0.07 K. 

In conclusion, we succeeded in detecting the zero-mode and its spin- and valley-split levels in the multilayered MDF system $\alpha$-(BEDT-TTF)$_2$I$_3$ probed by inter-layer magnetoresistance measurements. The effective Coulomb interaction plays an important role in intensifying the spin splitting of zero-mode carriers. Hence, the effective g-factor exceeds 2 and depends strongly on the magnetic field as $g^* \propto 1/\sqrt{B}$. The twofold valley degeneracy, on the other hand, is broken in the magnetic field that satisfies $\sqrt{\pi/2}I > \Gamma$. At 7 T, valley splitting abruptly emerges at about 2 K and increases up to approximately 17 K at 0.07 K. This system is very pure and therefore, we could detect the effects of Coulomb interaction and its associated valley splitting in the magnetic field, which is much lower than that of graphene. 

Recently, the multilayered MDF system was also realized in one of the parent compounds of the oxypnictide superconductors \cite{rf:18}. This MDF system, however, is not simple because this compound has multiband structure and therefore some Fermi surfaces. In this sense, $\alpha$-(BEDT-TTF)$_2$I$_3$ provides a suitable testing ground for the transport of the multilayered MDF system. According to the theory by Shon and Ando, the mono-layered MDF system show a universal conductance, $\sigma_{xx} \sim 2e^2/\pi^2\hbar$, which is independent of the magnetic field, the temperature and the scattering at Dirac point \cite{rf:19}. In the multilayered MDF system, however, the tunneling between layers reflects the broadening of zero-mode and it loses the validity of universal conductance in the magnetic field. Realistic theory reproduced well our anomalous in-plane magnetoresistance \cite{rf:7, rf:17}. Further investigation will lead us to new phenomena characterized by the multilayered MDF system.

We thank Dr. A. Kobayashi, Prof. Y. Suzumura, Dr. T. Morinari, Prof. T. Tohyama and Prof. T. Osada for valuable discussions. This work was supported by KAKENHI (Nos. 22540379 and 22224006).


\end{document}